\documentclass[aps,showpacs,10p]{revtex4-1}
\usepackage{graphicx}% Include figure files
\usepackage{dcolumn}% Align table columns on decimal point
\usepackage{bm}% bold math
\usepackage{hyperref}% add hypertext capabilities
\usepackage{amssymb}%\usepackage[mathlines]{lineno}% Enable numbering of text and display math
\usepackage{hyperref}% add hypertext capabilities
\RequirePackage[normalem]{ulem} %DIF PREAMBLE
\RequirePackage{color}\definecolor{RED}{rgb}{1,0,0}\definecolor{BLUE}{rgb}{0,0,1} %DIF PREAMBLE
\begin{document}
%\date{}
%\unitlength=1.00mm \special{em:linewidth 0.4pt}
%\linethickness{0.4pt}
%\thispagestyle{empty}
\newcommand{\be}{\begin{equation}}
\newcommand{\ee}{\end{equation}}
\newcommand{\ba}{\begin{eqnarray}}
\newcommand{\ea}{\end{eqnarray}}
\newcommand{\Gam}{\Gamma[\varphi]}
\newcommand{\Gamm}{\Gamma[\varphi,\Theta]}
\thispagestyle{empty}

\date{\today}

\title{ The von Neumann entanglement entropy for Wigner-crystal states in  one dimensional $N$-particle
systems}

\author{ Przemys\l aw Ko\'scik, Institute of Physics,  Jan Kochanowski University\\
ul. \'Swi\c{e}tokrzyska 15, 25-406 Kielce, Poland }

\begin{abstract}
We study  one-dimensional systems of $N$ particles in a
one-dimensional  harmonic trap with an inverse power law
interaction $\sim|x|^{-d}$. Within the framework of
the harmonic approximation we derive, in the strong interaction limit,
the Schmidt decomposition of the one-particle reduced density matrix
and investigate the nature of the degeneracy appearing in its spectrum.
Furthermore,  the ground-state asymptotic occupancies and their natural
orbitals are derived in closed analytic form, which enables
 their easy determination for a wide range of values of $N$. A closed form asymptotic expression for the
von Neumann entanglement entropy is also provided and its dependence
on $N$ is discussed for the systems with  $d=1$ (charged particles)
and with $d=3$ (dipolar particles).
\end{abstract}

 \maketitle
\section{Introduction}\label{1}

Within the last few years, systems of confined particles have
attracted considerable research attention because of the development
of new technologies and experimental techniques   that have opened
up perspectives for their experimental fabrication. Besides
the systems with short-range interactions, such as the Tonks--Girardeau
(TG) gases \cite{fer}, systems  with particles exhibiting
long-range interactions have also received a lot of attention in
recent years. Among them, the best known are the systems  of
spatially confined particles with a  Coulomb \cite{dot,ions} interaction or a
dipole--dipole \cite{ddi} interparticle interaction. Most
recently, there has been  a remarkable increase of interest
in the study of quantum entanglement, as entangled systems play an
essential role in  quantum information technology \cite{com}.
  In particular, a few
attempts have been made recently towards understanding the
entanglement in systems of interacting particles.
 For example, some light has been shed  on
entanglement both in quantum dot systems
\cite{mozano,1,2,7.2,3,9.2,kos2,ce,ghjj,maj} and
 in systems of harmonically interacting particles in a harmonic trap (the so-called Moshinsky
 atom) \cite{ww2,ww8,ww9,yanes,3.1,nagy,nagy2,nagy3}.
 Special attention has also been paid to the study of
entanglement in the helium atoms and helium ions
\cite{ww6,beneti,lin,ho,ho1,ho2,ho3,hokos,h4}. The effect of the
confinement on the entanglement in the ground state of the helium
atom has also been discussed in the literature \cite{cong}. For
 details on recent developments in entanglement studies of quantum composite systems,  see   \cite{tichy} for an
overview. However,  there has been relatively little research   so
far on entanglement in systems with more than two particles. The
reason for this lies in the fact that in most cases the
determination of the wavefunction of a few-body state requires
performing numerical calculations, which is a major problem in
general. According to our best knowledge, the only $N$ particle
system  of which the entanglement properties have been fully
explored so far is the Moshinsky model system \cite{ww8,ww9}

  The model on which we focus here
is a system composed of $N$ particles interacting via a long-range
inverse power-law potential, which are confined by an external
one-dimensional (1D) harmonic potential
 \be H=-{1\over
2}\sum_{i=1}^N{\partial_{x_{i}}^2}+V^{g}(x_{1},x_{2},...,x_{N}),\label{hamk}\ee
\be V^{g}(x_{1},x_{2},...,x_{N})={1\over 2}\sum_{i=1}^{N} x_{i}^2
 +\sum_{i>j=1}^{N} {g\over |x_{i}-x_{j}|^d},  d>0\label{pot}. \ee
Due to the singular behaviour of the interaction potential
$|x_{i}-x_{j}|^{-d}$ at $x_{i}=x_{j}$, the wavefunction of bosons
($B$) satisfies $\psi_{B}(x_{1},...,x_{N})=0$ if $x_{i}=x_{j}$, for
any $g\neq 0$ \cite{fer}. For example, when $d=1$ the system
 serves as a prototype for ions in electromagnetic traps
and when $d=3$, as a model of confined dipolar particles. Here we
address mainly the limit as $g\rightarrow\infty$ in which the
perfect linear Wigner molecule is formed
  \cite{wigner,wigner1,wigner2}, and the harmonic
approximation (HA)  is valid \cite{9.2,ghjj,ant,ant1}. Recently,
within the framework of this approximation,
 we have  derived when $N=3$ and $d=1$ an explicit integral representation for the asymptotic occupancies  and their natural orbitals
\cite{9.2}. In the present paper, we  extend the results of
\cite{9.2} and provide  the Schmidt expansions of the asymptotic
one-particle reduced density matrix (RDM) in  the general case of
$N$ particles, investigating thereby the nature of the degeneracy
appearing in its spectrum. Here we go even further and derive for
the ground state of (\ref{hamk}) closed form expressions for the
asymptotic occupancies and their natural orbitals, which enables
their easy determination for
 a wide range of values of $N$ without the solution of the eigenvalue problem with the RDM. Moreover, we derive an asymptotic
closed form expression for the von Neumann (vN) entropy and analyze
its dependence on $N$ for the systems with $d=1$ (charged particles)
and with $d=3$ (dipolar particles). Additionally, in the first case,
we address the question of how the changes in the parameter $g$
affect the entanglement.

The structure of this letter is as follows. In Section \ref{sed}, we
derive, within the framework of the harmonic approximation, an
asymptotic Schmidt decomposition of the RDM and provide  closed-form
asymptotic expressions for the ground state occupancies and the  vN
entropy. Section \ref{result} focuses on these results and some
concluding remarks are made in Section \ref{4}.

\section{The strong interaction limit}\label{sed}

\subsection{Harmonic approximation}\label{poloooo}

 The potential $V^{g}$ given by Eq. (\ref{pot}) attains its
minimum at  points
$\vec{r}_{min}=(x^{c}_{1},x^{c}_{2},...,x^{c}_{N})$ which are
determined by a
 set of  equations $\partial _{x_{k}} V^{g}(x_{1},...,x_{N})=0$, $k=1,...,N$.
As is easy to see, the potential (\ref{pot}) scales as \be
V^{g}(\beta_{1} g^{{1\over 2+d}},...,\beta_{N} g^{{1\over
2+d}})=g^{{2\over 2+d}}V^{g=1}(\beta_{1},...,\beta_{N}),\label{f}\ee
 which means that the equilibrium
positions of the particles $x_{i}^c$ have the form
$x_{i}^{c}=\beta_{i} g^{{1\over 2+d}}$, where $\beta_{i}$ are the
solutions of $\partial _{\beta_{k}}
V^{g=1}(\beta_{1},...,\beta_{N})=0$, $k=1,...,N$. The values of
$\beta_{i}$ have simple analytic forms only in the cases $N=2,3$.
For larger $N$, we find them numerically.

If   $g$ is large enough,  the particles  crystallize around their
classical equilibrium positions and  the HA can be applied
\cite{ant,ant1}.  From here on we refer to the point
 $\vec{r}_{min}$ with
$x^{c}_{1}<x^{c}_{2}<...<x^{c}_{N}$  without loss of generality. We
denote it by $\vec{r}_{min}^{(0)}$. In this case
$x_{i}^{c}=-x_{N-i+1}^{c}$ ($x_{{N+1\over 2}}^{c}=0$ if $N$ is odd
), $i=1,2,3,...,{P\over 2}$, where $P = N$ if $N$ is even but $N-1$
for odd values of $N$ \cite{ant}.  As is well known, within the HA,
the
 problem
 is  equivalent to a set of
$N$ uncoupled oscillators \cite{ant,ant1}, \be
H^{\mbox{HA}}=\sum_{i=1}^{N}{-{1\over 2
}d^2_{\zeta_{i}}}+{\omega_{i}^2 \zeta_{i}^2 \over 2},\label{pl}\ee
 where
   the values of  $\omega_{i}^2$ are
  the eigenvalues of the Hessian matrix $\textbf{H}=
[\frac{\partial^2 V^{g}}{\partial x_m\partial
x_k}|_{\{\vec{r}_{min}^{(0)}\}}]_{N\times N}$, and
$\vec{\zeta}=(\zeta_{1},\zeta_{2},...,\zeta_{N})$ are the so-called
normal modes given by $\vec{\zeta}=\textbf{U} \vec{\textbf{Z}}$,
where $\vec{\textbf{Z}}=(z_{1},z_{2},...,z_{N})$,
$z_{i}=x_{i}-x^{c}_{i}$, and $\textbf{U}$ is the matrix of the
corresponding eigenvectors of $\textbf{H}$.

The transformation $x_{i}=\beta_{i} g^{{1\over 2+d}}$ yields
$$\frac{\partial^2 V^{g}(x_{1},...,x_{N})}{\partial x_m\partial
x_k}=g^{-{2\over 2+d}}\frac{\partial^2 V^{g}(\beta_{1} g^{{1\over
2+d}},...,\beta_{N} g^{{1\over 2+d}})}{\partial \beta_m\partial
\beta_k},$$ so that, due to (\ref{f}), we can readily  conclude that
the Hessian matrix does not depend on $g$ but only on
$\{\beta_{i}\}$.

For transparency of presentation we concentrate on analyzing the
spatially symmetric ($+$) and antisymmetric ($-$) ground states.
 The normalized eigenfunction of (\ref{pl}) with   lowest
energy  is given by \be \psi
 (z_{1},z_{2},...,z_{N})=\prod_{i=1}^N
({{\omega_{i}}\over \pi})^{1\over 4}e^{-{\omega_{i}
\zeta^{2}_{i}(z_{1},z_{2},...,z_{N})\over 2}},\label{pp}\ee
 and provides
  an approximation to   wavefunctions of (\ref{hamk})
 only around $\vec{r}_{min}^{(0)}$.
In terms of  (\ref{pp}),  approximations to the spatially symmetric
($+$) and antisymmetric ($-$) ground-state wavefunctions of
(\ref{hamk}) can be  written in  forms convenient for further
analysis as  \cite{9.2,maj}
\begin{eqnarray}\Psi^{(\pm)}(x_{1},x_{2},...,x_{N})= \aleph^{(\pm)} \sum_{p}sgn(p)\psi(x_{p(1)}-x^{c}_{1},x_{p(2)}-x_{2}^{c},...,x_{p(N)}-x_{N}^{c})
\label{pps},\end{eqnarray}  and
 the sum goes over all permutations, and  $sgn(p)$ is $1$ for bosons
while for fermions $sgn(p)$ is $1$ for even permutations and $-1$  for odd
ones. The
 normalization factor $\aleph^{(\pm)}$ tends  to $N!^{-{1/ 2}}$ as $g\rightarrow\infty$ both in the bosonic and fermionic
 case, which is a consequence of  the fact that the classical distance between the classical
equilibrium positions of any pair of particles
 tends to infinity as $g\rightarrow\infty$, that is, $\Delta_{ij}=|x_{i}^{c}-x_{j}^{c}|\rightarrow\infty$
 for any $i\neq j$ (note that
  the
integral overlap between  any pair of  functions
$\psi(x_{p(1)}-x^{c}_{1},x_{p(2)}-x_{2}^{c},...,x_{p(N)}-x_{N}^{c})$
with different permutations of the coordinates vanishes). For the
same reasons, the RDM  of (\ref{pps}),
\begin{eqnarray} \rho^{(\pm)}(x,y)=\int_{\Re^{N-1}}(\prod_{k=2}^N
dx_{k})
\Psi^{(\pm)}(x,x_{2},...,x_{N})\Psi^{(\pm)}(y,x_{2},...,x_{N}),\label{rdm}\end{eqnarray}
reduces in  the $g \rightarrow \infty$ limit  to
 \be
\rho^{g\rightharpoonup\infty}(x,y)=\sum_{i=1}^{N}{\rho}_{i}(x,y),\label{ex}\ee
 with
\begin{eqnarray}
\rho_{i}(x,y)=\nonumber\\{1\over N}\int_{\Re^{N-1}}(\prod_{ k\neq i}
d\mu_{k})\psi(\mu_{1},...,\mu_{i-1},x-x^{c}_{i},\mu_{i+1},...,\mu_{N})\nonumber\\\times\psi(\mu_{1},...,\mu_{i-1},y-x^{c}_{i},\mu_{i+1},...,\mu_{N}).\label{partial}\end{eqnarray}
 Note that (\ref{ex}) is
properly normalized, namely $Tr
\rho^{g\rightharpoonup\infty}=\int_{\Re}\rho^{g\rightharpoonup\infty}(x,x)dx=1$.
The asymptotic behaviour for the RDM is thus the same for the case
of bosons and fermions. It should be stressed that  Eqs.
(\ref{ex})--(\ref{partial})  have already appeared before in
\cite{maj}. Moreover, in \cite{maj},
 an asymptotic   linear  entropy $\mbox{L}$ of the ground state of the system (\ref{hamk}) with $d=1$ was computed   directly  from the integral
  representation $\mbox{L}=1-\int_{_{\Re^{2}}} [\rho^{g\rightharpoonup\infty}(x,y)]^2 dxdy$ (for $N$ up to
  $N=10$).
 The present paper goes substantially further as it  discusses the Schmidt
decomposition of (\ref{ex}) and provides closed form expressions for
the occupancies, natural orbitals, and the vN entropy as well.
Moreover, it  provides the results  for the entanglement both in the
  $d=1$ and  $d=3$ case.

\subsection{Schmidt decomposition  of the RDM}
To start with, let us  note  that the equilibrium positions in Eq.
(\ref{partial}) can be eliminated by translating the coordinates
 by $x\mapsto \tilde{x}+x^{c}_{i}$, $y\mapsto
\tilde{y}+x^{c}_{i}$,
${\rho}_{i}(x,y)\mapsto{\tilde{\rho}}_{i}(\tilde{x},\tilde{y})$.
Being normalizable, real, and symmetric under permutations of
coordinates, the function ${\tilde{\rho}}_{i}$ has a Schmidt
decomposition in the form  \cite{vn1}\be
{\tilde{\rho}}_{i}(\tilde{x},\tilde{y})=\sum_{l=0}^{\infty}\lambda_{l}^{(i)}v^{(i)}_{l}(\tilde{x})v^{(i)}_{l}(\tilde{y}),\label{kkl}\ee
$\langle v^{(i)}_{l}|v^{(i)}_{k}\rangle=\delta_{lk}$, where
$v^{(i)}_{l}$ and  $\lambda_{l}^{(i)}$ satisfy the integral
eigensystem  \be
\int_{-\infty}^{\infty}\tilde{\rho}_{i}(\tilde{x},\tilde{y})v_{l}^{(i)}(\tilde{y})
d\tilde{y}=\lambda_{l}^{(i)}v_{l}^{(i)}(\tilde{x}).\label{rrd1}\ee
By changing the variables back in (\ref{kkl}), namely by
$\tilde{x}\mapsto x-x^{c}_{i}$, $\tilde{y}\mapsto y-x^{c}_{i}$, one
gets \be
{\rho}_{i}(x,y)=\sum_{l=0}^{\infty}\lambda_{l}^{(i)}v^{(i)}_{l}(x-x^{c}_{i})v^{(i)}_{l}(y-x^{c}_{i}),\ee
and substituting the above into (\ref{ex}), we arrive at
\begin{eqnarray} {\rho}^{g\rightharpoonup\infty}(x,y)=\sum_{i=1}^N\sum_{l=0}^{\infty}\lambda_{l}^{(i)}
v_{l}^{(i)}(x-x^{c}_{i})v_{l}^{(i)}(y-x^{c}_{i}).\label{rdml1}\end{eqnarray}
One can easily infer  that in the limit as $g\rightarrow\infty$,
where $\Delta_{ij}\rightarrow \infty$ ($i\neq j$),  the integral
overlap $\langle
v_{k}^{(i)}(x-x^{c}_{i})|v_{l}^{(j)}(x-x^{c}_{j})\rangle$
 vanishes for any
$k, l$ as long as $i\neq j$. Hence, bearing in mind that
 $\langle
v^{(i)}_{l}(x-x^c_{i})|v^{(i)}_{k}(x-x^{c}_{i})\rangle=\delta_{lk}$,
we  can conclude  that the family
$\{v^{(i)}_{l}(x-x^{c}_{i})\}_{i=1,l=0}^{N,\infty}$ forms
 an orthonormal set as $g\rightarrow\infty$. In this limit the
 expansion (\ref{rdml1})
is therefore  nothing else but the Schmidt decomposition of the
asymptotic RDM (\ref{ex}), which means  $v^{(i)}_{l}(x-x^{c}_{i})$ are its eigenvectors (natural orbitals)
 and   $\lambda_{l}^{(i)}$ are its eigenvalues (occupancies).

The normal-mode coordinates have the form
$\zeta_{i}=U_{i,1}z_{1}+U_{i,2}z_{2}+...+U_{i,N}z_{N}$, where
the $U_{i,j}$, being the elements of the matrix $\mbox{U}$, satisfy
  $U_{i,j}=-U_{i,N-j+1}$ (here $U_{i,{(N+1)/
2}}=0$ if $N$ is odd) or $U_{i,j}=U_{i,N-j+1}$, for all indices $j$
at fixed $i$ \cite{ant}. One can readily infer from the above that
${\tilde{\rho}}_{i}={\tilde{\rho}}_{N-i+1}$ and, as a result, we
have
  $v_{l}^{(i)}(\tilde{x})=v_{l}^{(N-i+1)}(\tilde{x})$ and
$\lambda_{l}^{(i)}=\lambda_{l}^{(N-i+1)}$. Thus, conservation of
probability  gives
($M_{i}=\sum_{l=0}^{\infty}\lambda_{l}^{(i)}$): $ 2\sum_{i=1}^{N/
 2}M_{i}=1$
and $ 2\sum_{i=1}^{{(N-1)/
 2}}M_{i}+M_{{(N+1)/
 2}}=1$, for
even and odd values of $N$,
 respectively.
 All the asymptotic occupancies except those corresponding to
${\rho}_{{(N+1)/
 2}}$ ($N$ odd) are thus  two-fold degenerate.

Since the asymptotic RDM possesses a degenerate spectrum, its
Schmidt decomposition fails to be unique \cite{uniq}. For the sake
of completeness, we derive below another form of the Schmidt
decomposition of the asymptotic RDM. Accordingly, for each
double-point of degeneracy,
$\lambda_{l}^{(i)}=\lambda_{l}^{(N-i+1)}$, we  define from the
corresponding orbitals $v_{l}^{(i)}$ and $v_{l}^{(N-i+1)}$, the new
ones as \be
\eta_{l}^{(i)}(z)={v_{l}^{(i)}(z-x_{i}^{c})+v_{l}^{(N-i+1)}(z-x_{N-i+1}^{c})\over
\sqrt{2}},\label{df}\ee
 and
\be
\tau_{l}^{(i)}(z)={v_{l}^{(i)}(z-x_{i}^{c})-v_{l}^{(N-i+1)}(z-x_{N-i+1}^{c})\over
\sqrt{2}},\label{df1}\ee that fulfill $\langle
\eta_{l}^{(i)}|\tau_{l}^{(i)}\rangle=0$. As is easy to check in terms
of (\ref{df})--(\ref{df1}), Eq. (\ref{rdml1}) can be rearranged to
\begin{eqnarray}
\rho^{g\rightarrow\infty}(x,y)=\sum_{i=1}^{{N\over
2}}\sum_{l=0}^{\infty}\lambda_{l}^{(i)}[\eta_{l}^{(i)}(x)\eta_{l}^{(i)}(y)+\tau_{l}^{(i)}(x)\tau_{l}^{(i)}(y)],\label{sd}\end{eqnarray}

\begin{eqnarray}
\rho^{g\rightarrow\infty}(x,y)=\sum_{l=0}^{\infty}\lambda_{l}^{({N+1\over
2})}v_{l}^{({N+1\over 2})}(x)v_{l}^{({N+1\over
2})}(y)+\sum_{i=1}^{{N-1\over
2}}\sum_{l=0}^{\infty}\lambda_{l}^{(i)}[\eta_{l}^{(i)}(x)\eta_{l}^{(i)}(y)+\tau_{l}^{(i)}(x)\tau_{l}^{(i)}(y)],\label{sd1}\end{eqnarray}
for  even and odd values of $N$, respectively. In the limit as
$g\rightarrow \infty$,
 we have
$\langle\eta_{l}^{(i)}|\eta_{k}^{(j)}\rangle=\delta_{lk}\delta_{ij}$,
$\langle\tau_{l}^{(i)}|\tau_{k}^{(j)}\rangle=\delta_{lk}\delta_{ij}$
and the integral overlaps $\langle
\eta_{l}^{(i)}|\tau_{k}^{(j)}\rangle$, $\langle v_{l}^{({N+1\over
2})}|\eta_{k}^{(j)}\rangle$, $\langle v_{l}^{({N+1\over
2})}|\tau_{k}^{(j)}\rangle$,  vanish for any $l,k$ and $i,j$,
($i,j=1,2,3,...,{P\over 2}$ with $P$ as before).
  We
therefore can recognize (\ref{sd})--(\ref{sd1}) as a Schmidt form of
the asymptotic RDM different from that given by (\ref{rdml1}).

\subsection{Occupancies}
 By
inspection, the integrals (\ref{partial}) can be performed
explicitly: \be \tilde{\rho}_{i}(\tilde{x},\tilde{y})= A_{i}
e^{-a_{i}(\tilde{x}^2+\tilde{y}^2)-b_{i}
\tilde{x}\tilde{y}},a_{i}>0, b_{i}<0, \label{polo}\ee where,
however, due to the reasons mentioned at the beginning of the
section \ref{poloooo}, the coefficients $A_{i}$, $a_{i}$, $b_{i}$
can be found in closed analytic forms only in the cases $N=2$ and
$N=3$. Nonetheless, it turns out that
 even  in the former case they have quite complicated forms: $$A_{i}={{\sqrt[4]{d+2}}\over{
\sqrt{2 \pi(\sqrt{d+2}+1)}}},$$ $$a_{i}={{d+6 \sqrt{d+2}+3}\over
{8(1 +\sqrt{d+2}})},$$ $$b_{i}=-{d-2 \sqrt{d+2}+3\over 4(1+
\sqrt{d+2})},$$ ($i=1,2$). We have noted that the  Schmidt
decomposition of the function (\ref{polo}), thereby the asymptotic
occupancies $\lambda_{l}^{(i)}$ and their natural orbitals
$v_{l}^{(i)}$, can be found in closed form. Here we proceed
similarly to \cite{nagy}, wherein  the occupancies of the
analytically solvable two-particle Moshinsky model were derived by
the use of Mehler's formula:
\begin{eqnarray}e^{- (u^2+v^2){y^2\over 1-y^2} +uv {2y\over 1-y^2}}
=\sum_{l=0}^{\infty}\sqrt{1-y^2}({y\over 2})^l {\mbox{H}(l;u)
\mbox{H}(l;v)\over l!},\label{meh}\end{eqnarray} where ${H}(l;.)$ is
the $l$th order Hermite polynomial. Indeed, by matching Eq.
(\ref{polo}) with (\ref{meh}), one arrives,  after performing some
tedious algebra, at

\begin{eqnarray} \tilde{\rho}_{i}(\tilde{x},\tilde{y})=A_{i}
e^{-a_{i}(\tilde{x}^2+\tilde{y}^2)-b_{i}
\tilde{x}\tilde{y}}=\sum_{l=0}^{\infty}\lambda_{l}^{(i)}v_{l}^{(i)}(\tilde{x})v_{l}^{(i)}(\tilde{y}),\end{eqnarray}
with \be v_{l}^{(i)}(\tilde{x})={w_{i}^{1\over 4}\over \pi^{1\over
4}\sqrt{2^l l!}}e^{-{1\over 2}w_{i}
\tilde{x}^2}\mbox{H}(l;\sqrt{w_{i}} \tilde{x}),\label{nat}\ee and
\be \lambda_{l}^{(i)}=A_{i}\sqrt{{\pi(1-y_{i}^2)\over
w_{i}}}y_{i}^{l},\label{ocup}\ee where
\begin{eqnarray}w_{i}=\sqrt{4a_{i}^2-b_{i}^2},\nonumber\end{eqnarray}
\begin{eqnarray}y_{i}={\sqrt{2a_{i}-b_{i}}-\sqrt{2a_{i}+b_{i}}\over
\sqrt{2a_{i}-b_{i}}+\sqrt{2a_{i}+b_{i}}}.\nonumber\end{eqnarray} We
found that the values of two lowest asymptotic occupancies computed
by means of (\ref{ocup}) for $N=3,d=1$:
 $\lambda_{0}^{(1)}\cong0.324905$,
$\lambda_{0}^{(2)}\cong0.319336$ are in agreement with  the values
of \cite{9.2} ($0.3249$, $0.3193$), where they were determined by the
numerical solution of (\ref{rrd1}),
  which confirms the correctness of (\ref{nat})--(\ref{ocup}).

\subsection{The  von Neumann entropy}\label{asymvn} The entanglement   vN entropy is given by \cite{vn1}
\be \mbox{S}=-\mbox{Tr}[\rho \mbox{log}_{2}\rho],\ee and in terms of
the occupancies, it takes the form
$\mbox{S}=-\sum_{l}\lambda_{l}\mbox{log}_{2}\lambda_{l}$.  In view
of (\ref{sd})--(\ref{sd1}), it follows that the asymptotic vN
entropy can be decomposed for even and odd values of $N$  as \be
\mbox{S}^{g\rightarrow\infty}=2\sum_{i=1}^{N\over
 2}S_{i},\label{ee}\ee
and \be \mbox{S}^{g\rightarrow\infty}=2\sum_{i=1}^{{N-1\over
 2}}S_{i}+S_{{N+1\over
 2}},\label{ee1}\ee
 respectively, where the components
$S_{i}$ are given by  $S_{i}=-\sum
\lambda_{l}^{(i)}\mbox{log}_{2}\lambda_{l}^{(i)}$. Next, using the
analytical formula obtained by us for $\lambda_{l}^{(i)}$, Eq.
(\ref{ocup}), we attempt to derive a closed form expression for
$S_{i}$. We found that for this purpose   it is convenient to
rewrite $S_{i}$ as \be S_{i}=-\lim_{q\rightarrow1}\sum_{l=0}^\infty
[\lambda_{l}^{(i)}]^q {d\over dq}\mbox{log}_{2}(\sum_{l=0}^\infty
[\lambda_{l}^{(i)}]^q), \label{ent}\ee which can be easily verified
by referring to the derivative formulas $(\mbox{log}_{a}
x)^{'}=[1/x\mbox{ln}a],(a^x)^{'}=a^x \mbox{ln}a$. By performing the
summation  in (\ref{ent}), we obtain
 \be
\sum_{l=0}^\infty [\lambda_{l}^{(i)}]^q=\frac{\pi ^{q/2}
\left({A_{i}
   \sqrt{1-y_{i}^2\over w_{i}}}{}\right)^q}{1-y_{i}^q}.\label{ll}\ee
Substituting this  into (\ref{ent}) and  then taking the limit as
$q\rightarrow1$, we get  \be S_{i}=-\frac{ A_{i}
\sqrt{\frac{\pi(y_{i}+1)}{w_{i}}} \mbox{ln}\left(  y_{i}^{2y_{i}}
\left(\frac{\pi A_{i}^2
   \left(1-y_{i}^2\right)}{w_{i}}\right)^{1-y_{i}}\right)}{(1-y_{i})^{3/2} \mbox{ln} (4)}.\label{kl3} \ee

\section{Numerical results}\label{result}
 In our research for finding
the ground-state approximate wavefunction of (\ref{hamk}) for finite
values of $g$, we apply a simple trial wave function given by a
Jastrow-type
 wavefunction:
 \be
\chi(x_{1},x_{2},...,x_{N})\propto\prod_{k=1}^Ne^{-{x_{k}^2\over
2}}\prod_{i>j}^{N}f(\alpha{x_{i}-x_{j}\over \sqrt{2}}),\label{kk}\ee
with \be f(x)=e^{{x^2/ 2}}\phi( x),\label{rrrf}\ee where $\phi$ is a
relative motion wavefunction for the ground state of the
two-particle system and $\alpha$ is a variational parameter that is
optimized in order to minimize the expectation value of  the energy.
Such a form of $f$ was recently suggested in \cite{cremon} and its
applicability to the case of charged particles has been
demonstrated. We recall at this point that in the limit as
$g\rightarrow 0$, the system (\ref{hamk}) forms a TG gas and its
exact ground-state wavefunction is given by (\ref{kk}) with $f=|x|$
\cite{fer}.
 In
the case of $N=2$, the Hamiltonian (\ref{hamk}) separates in terms
of
$$ x={{x}_{2}-{x}_{1}\over \sqrt{2}}, X={x_{1}+x_{2}\over
\sqrt{2}},$$ into $H= H^{x} +H^{X}$, where  $H^{X}=-{1/
2}{d_{X}^2}+{1/ 2} X^2$
 is exactly solvable and $H^{x}$ is given by
\be H^{x}=-{1\over 2} d_{x}^{2} +{1\over 2} x^2+{g\over 2^{d\over
2}|x|^d}.\label{oko}\ee  Because the  interaction $ |x|^{-d}$
$(d>0)$ diverges when $x\rightarrow 0$,  the ground-state even
wavefunction $\phi^{(+)}$ of (\ref{oko}) is given by
$\phi^{(+)}=|\phi^{(-)}|$, where  $\phi^{(-)}$ ($\phi^{(-)}(0)=0$)
is the lowest energy odd wavefunction. Note that the function
(\ref{kk}) with $\phi=|\phi^{(-)}|$ is properly defined  for the 1D
bosons, i.e., it is symmetric and $\chi(x_{1},x_{2},...,x_{N})=0$ if
$x_{i}=x_{j}$.

\begin{table}[h]
\begin{center}
\begin{tabular}{llllll}
\hline & $N$
&  $g={1\over 2}$&$g=2$ &$g=7$ \\
\hline

Ref.\cite{kos2}&3 &${0.546 } $ &${0.609 }$ &${0.682}    $ \\
L &&$0.551(0.9) $ &$0.616(0.88) $ &$0.679(0.87)  $ \\
Ref.\cite{kos2} &4&${0.628 } $ &${0.685 }$ &${0.765}    $ \\
L &&$0.636(0.833) $ &$0.694(0.8) $ &$ 0.758 (0.78) $ \\
\end{tabular}
\caption{\label{tab:table5p1p} Linear entropies obtained as
discussed in the text are compared with those obtained in
\cite{kos2} (Fig. 6 in it) by means of the CI method. The optimal
values of $\alpha$  are placed in brackets. Here in each case
$\Delta y=0.25$ was used. }

\end{center}
\vspace{-0.6cm}
\end{table}

 The occupancies and their natural orbitals are determined by the following  integral eigenequation $
\int_{-\infty}^{\infty}{\rho}(x,y)v_{s}(y) dy=\lambda_{s}v_{s}(x)$,
which can be turned into an algebraic problem by discretizing the
variables $x$ and $y$ with equal subintervals of length  $\Delta y$
 (see, for example, \cite{ghjj}). Thus, the eigenvalues of the matrix
$\mbox{B}=[\Delta y \rho(m_{i},m_{ j})]_{K\times K}$, where
$m_{i}=-c+\bigtriangleup y i$, $\bigtriangleup y=2c/(K-1)$,
$i,j=0,...,K-1$, provide approximations to the $K$ largest
occupancies. The method produces reasonable approximations to the
lowest  occupancies if only the value of $\Delta y$ is small enough
and the value of $2c$ is at least as large as the side of a square in
which $\rho(x,y)$ is mainly confined. A good criterion for
convergence  is the closeness of the sum of approximate occupancies to
the theoretical value $1$.
 In this paper,
both for the optimization of
  (\ref{kk}) with respect to $\alpha$ and for the determination of
$\mbox{B}$,  Monte Carlo techniques are used
(for $N>3$). An exception is the case of $g\rightarrow 0$
(the TG limit), where an effective algorithm presented  by Pezer and
Buljan \cite{dd}  is used to determine the occupancies.

 In order to gain insight into the effectiveness
of the ansatz (\ref{kk}), we first determine the occupancies
$\lambda_{s}$ of the ground states of three- and four- particle
systems with $g=0.5,2,7$ ($d=1$) and assess their accuracy by
comparing the linear entropies $\mbox{L}=1-\sum_{s}\lambda_{s}^2$
with
  the numerical results of
 \cite{kos2}, wherein the corresponding values of $\mbox{L}$ were determined by the CI method. In each case, we
solve Eq. (\ref{oko}) for the corresponding function  $\phi^{(-)}$
by the Rayleigh--Ritz method with a set of the ten lowest odd
eigenfunctions of 1D harmonic oscillator.
 Our numerical results, together
with the ones of \cite{kos2} are summarized
 in Table \ref{tab:table5p1p},   where, for the sake of completeness,  we also give  the optimal
values of $\alpha$. It is apparent   from the results  that the
ansatz (\ref{kk}) gives a reasonable estimate of the linear entropy,
which allows us to expect that it also provides  good approximations
to the
 true values of the occupancies.

\begin{figure}[h]
\begin{center}
\includegraphics[width=0.65\textwidth]{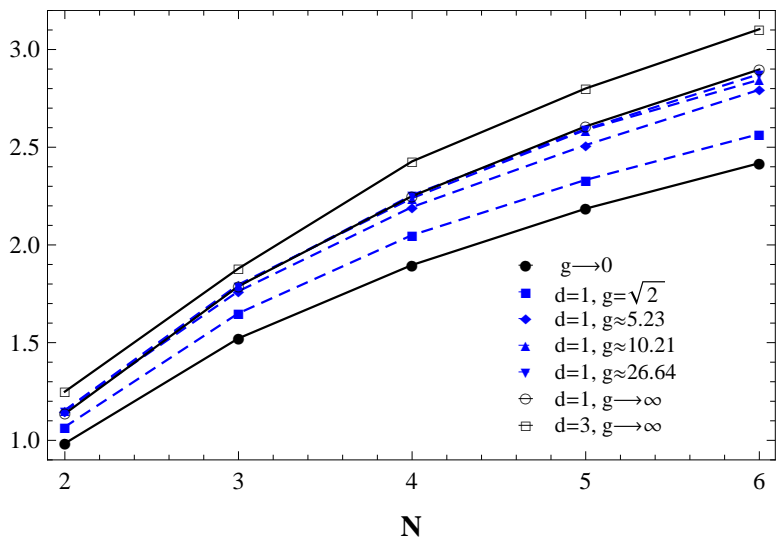}

\end{center}
\caption{\label{fffklolofghk3lg:beh} The vN entropy as a function of
$N$
 for the systems
with $d=1$ for $g\rightarrow0,g=\sqrt{2}, g\thickapprox 5.23,
g\thickapprox 10.21, g\thickapprox26.64, g\rightarrow\infty$, and
for the system with $d=3$ for $g\rightarrow\infty$.}
\end{figure}

Let us now move to  the point where we explore the effects of
changing the control parameters of the system (\ref{hamk}) on the
behaviour of the ground state vN
 entropy. As it comes down to  the
effect of $g$, we restrict our study
 to the case $d=1$. In order to make
our calculations as efficient as possible, we perform them for
some values of $g$ for which the functions $f$ can be derived in
closed exact forms (see Appendix). Here we consider the cases
 $g =
\sqrt{2}$ ($n=1$), $g\approx5.231$ $(n=3)$, $g\approx10.2176$
($n=5$),
 $g\approx$26.640 ($n=10$) where an especially  simple form of $f$ is obtained in the first case,
$f=|x|+x^2$.  Our numerical results for $\mbox{S}$, including the
limiting case of $g\rightarrow 0$, are  depicted in
 Fig.
\ref{fffklolofghk3lg:beh} together with the results  obtained   for
$\mbox{S}^{g\rightarrow\infty}$  by means  of (\ref{ee})--(\ref{ee1})
and (\ref{kl3}). Moreover, in order to gain some insight into how
the parameter $d$ influences the entanglement,  the variation of
$\mbox{S}^{g\rightarrow\infty}$  for  $d=3$  is also shown in this
figure. We can observe how $\mbox{S}$ converges to the values
predicted by the HA  as $g$ is increased, which confirms  both
the correctness of our theoretical derivations and the effectiveness of
(\ref{kk}). As can be seen, for all cases considered,
 the vN entropy
$\mbox{S}$ shows a monotonic increase with $N$. We conclude from the
results that the range of values of $g$ in which $\mbox{S}$ makes
its most rapid variation also increases with $N$. In other words,
the larger is  $N$, the larger is the value of $g$ at which
$\mbox{S}$ starts to approach its asymptotic value.
 For example, as one can infer from Fig. \ref{fffklolofghk3lg:beh}, the systems with $ N=4$ and with $N=6$ start to reach  their asymptotic
 behaviors practically when $g$ exceeds the values of about $g=10$ and $g=26$, respectively.

 We found  that if $N$ is relatively small and
$g\rightarrow\infty$, then only the occupancies with $l=0$, that is
$\lambda_{0}^{(i)}$, contribute considerably to the conservation of
the probability. More precisely, as we have verified,
 their sum  over  $i$ generally decreases as $N$ increases. For example, in the case
of $d=1$ considered in Fig.\ref{fffklolofghk3lg:beh},
 we found that it falls off from
$0.982$ at $N=2$ to $0.947$ at $N=6$. We have thus a qualitative
result that in the expansions of the RDM of   Wigner molecule state
(Eq. (\ref{rdml1}) and
 Eqs. (\ref{sd})--(\ref{sd1}))
 formed by a small $N$, only the terms with $l=0$ are substantial.
 Finally, by comparing
the asymptotic results obtained for  $d = 1$ and for
 $d = 3$, we arrive at the
conclusion that strongly interacting charged bosons are less
entangled than the strongly interacting dipolar ones.

\section{Summary}\label{4}
We studied the asymptotic entanglement properties of  systems
composed of $N$ particles trapped in a 1D harmonic potential with
the inverse power law interparticle interaction $\sim |x|^{-d}$. We
focused mainly on the strong interaction limit of
$g\rightarrow\infty$ in which  Wigner crystal states are formed.
Based on the harmonic approximation, we investigated the nature of
the degeneracy appearing in this limit in the spectrum of the
  RDM, by providing   its asymptotic Schmidt
expansions. Moreover, we obtained closed form expressions for the
ground-state asymptotic natural orbitals and their occupancies by
use of Mehler's formula. Furthermore, based on this finding, we also
derived an analytical expression for the corresponding asymptotic
von Neumann entropy and provided the results for its dependence on
$N$ for the systems  of charged ($d=1$) and  dipolar ($d=3$)
particles. It turned out that the Wigner molecule states formed by
the dipolar particles are more entangled than those formed by the
charged ones. Moreover, in the  case $d=1$, we carried out a
comprehensive study of the effect of changing both $N$ and $g$ on
the behavior of the von Neumann entropy. Our results showed that the
von Neumann entropy grows monotonically with $N$. Among other
things, it was found that the range of values of $g$ in which the
von Neumann entropy makes its most rapid variations tends to
increase with $N$.

\section{Appendix}

We found that for a countably infinite set of $g$ values, the
wavefunctions of the relative motion Schr\"{o}dinger equation
\begin{equation}
\left[-{1\over 2}{d^2\over d x^2}+{1\over 2} x^2
 + {g \over \sqrt{2} |x|} \right]\phi(x)=E_{rel}\phi(x),\label{EOM2111}
\end{equation}
can be derived in closed form. The solutions of the above equation,
in the even-parity case, which is all that we are interested in
here, may be represented by \be \phi^{(+)}(x)=|x|e^{-{x^2\over 2}}
\sum_{k=0}^{\infty}a_{k} |x|^k\label{ser},\ee where the coefficients
of $a_{k}$  satisfy the recurrence relation  \be (-1-2
E_{rel}+2k)a_{k-2}+\sqrt{2}g a_{k-1}-k(k+1)a_{k}=0,\ee where
$a_{0}\neq0$ and $a_{k}=0$, for $k<0$. The series (\ref{ser})
terminates after the $n$th term if, and only if, $E_{rel}= (3+2n)/2$
and $a_{n+1}=0$, which for a given $n$ determines the values of
$E_{rel}$ and $g$ at which closed form solutions of (\ref{EOM2111})
can be found. For example for $n=1$ we find $g = \sqrt{2}$, which
corresponds to $ \phi^{(+)}=e^{-{x^2\over 2}}(|x|+x^2)$ and
$E_{rel}=2.5$.

\section{Acknowledgments}

Thanks go to Rafa{\l} Rzeszutko for   his  comments on a draft of
this article.

\bibliography{aipsamp}

\end{document}